\documentclass{article}
\usepackage{spconf,amsmath,graphicx}

\usepackage[T1]{fontenc}
\usepackage{graphicx}
\usepackage{cite}
\usepackage{subfig}
\usepackage{xcolor}
\usepackage{subfig}
\usepackage{optidef}
\usepackage{amssymb}
\usepackage{amsmath}
\usepackage{array}
\usepackage{mathtools}
\usepackage{scalerel}
\usepackage{algorithm}
\usepackage{algorithmic}

\title{Using Intelligent Reflecting Surfaces for Rank Improvement \\ in MIMO Communications }

\name{\"Ozgecan \"Ozdogan, Emil Bj\"ornson, Erik G. Larsson\thanks{\newline\indent The paper was supported by ELLIIT and Swedish Research Council. 
		\newline \indent The authors are with the Department of Electrical Engineering (ISY), Link\"{o}ping University, SE-58183 Link\"{o}ping, Sweden \{ozgecan.ozdogan,emil.bjornson,erik.g.larsson\}@liu.se.}
}
\address{Department of Electrical Engineering (ISY), Link\"oping University, Sweden}

\begin{document}
\ninept


\maketitle

\begin{abstract}
An intelligent reflecting surface (IRS), consisting of reconfigurable metamaterials, can be used to partially control the radio environment and thereby bring new features to wireless communications. Previous works on IRS have particularly studied the range extension use case and under what circumstances the new technology can beat relays. In this paper, we study another use case that might have a larger impact on the channel capacity: rank improvement. One of the classical bottlenecks of point-to-point MIMO communications is that the capacity gains provided by spatial multiplexing are only large at high SNR, and high SNR channels are mainly appearing in line-of-sight (LoS) scenarios where the channel matrix has low rank and therefore does not support spatial multiplexing. We demonstrate how an IRS can be used and optimized in such scenarios to increase the rank of the channel matrix, leading to substantial capacity gains.
\end{abstract}


\section{Introduction}
An intelligent reflecting surface (IRS) is a thin two-dimensional metamaterial (i.e., engineered material) that can control and transform electromagnetic waves \cite{Tretyakov,Alu}. It has been demonstrated experimentally that metasurfaces can dynamically produce unusual scattering, polarization, and focusing properties to obtain desired radiation patterns \cite{Yang2016, Wan2016}.  Thanks to rapid recent development in lumped elements such as micro-electro-mechanical-systems (MEMS), varactors and PIN diodes, and tunable materials such as liquid crystals and graphene, metasurfaces with flexible functionalities can be successfully realized with low implementation cost and light weight\cite{Hum}. This has opened up exciting opportunities to use metasurfaces to solve problems in wireless communication research.

Metasurfaces are implemented as an array of discrete scattering elements. Each element (also known as a meta-atom or lattice) has the ability to introduce a phase shift to an incident wave. The change in the local surface phase is achieved by tuning the surface impedance which enables manipulation of the impinging wave. This operation creates phase discontinuities and requires abrupt phase changes over the surface. The IRSs obey the generalized Snell's law \cite{Yu333} and their discrete structure provide great design flexibility.

The metasurface technology has recently gained interest in wireless communications, under names such as IRS \cite{Wu2018},  large intelligent surface \cite{Nadeem}, and software-controlled metasurfaces \cite{Akyildiz, BJORNSON2019}. Several promising use cases of this technology, such as for range extension to users with obstructed direct links \cite{Wu2018, Nadeem,8741198, Bjornson2019a }, joint wireless information and power transmission in internet of things (IoT) networks \cite{Wu2019b}, physical layer security \cite{Guan, Yu2019}, unmanned air vehicle (UAV) communications \cite{Li2019} have been studied. Still, the IRS-aided wireless communication systems are new and remain largely unexplored.

\subsection{Relation to Prior Work}

The existing works on IRS have particularly focused on the range extension application.  In this use case, the IRS is deployed between the base station (BS) and user equipment (UE) and assists the communication. Generally speaking, the optimized selection of the phase of each discrete element in the IRS leads to phase alignment of  the direct channel (BS to UE) and scattered channel (BS to IRS to UE). If the direct path is totally blocked then the coherent phase alignment of the scattered path is the main goal. 

In particular, \cite{Wu2018} pointed out that an IRS has the capability of improving poorly-conditioned multi-user MIMO channels by adding controllable multipaths in cases where each user is equipped with a single antenna. IRS-aided point-to-point multiple data stream MIMO setups with Rician fading channels are studied in \cite{Ye2019}, \cite{Zhang2019a}. In \cite{Ye2019}, the direct path is assumed to be totally blocked. In \cite{Zhang2019a}, the authors propose an alternating optimization algorithm for capacity maximization.

 In this paper, we study the rank improvement ability of an IRS-aided single-user MIMO system while preserving the coherent phase alignment by optimizing the phase shifts. One of the classical bottlenecks of point-to-point MIMO communications is that the capacity gains provided by spatial multiplexing are only large at high SNR, and high SNR channels are mainly appearing in LoS scenarios where the channel matrix has low rank and therefore does not support spatial multiplexing \cite{1542408, 6509469}. Using a different setup than \cite{Zhang2019a}, we demonstrate how an IRS can be used and optimized in an LoS environment to increase the rank of the channel matrix, leading to substantial capacity gains. The classical waterfilling algorithm is adapted to perform power allocation between the data streams.

\section{System Model}
 We consider communication from a multiple-antenna BS to a multiple-antenna UE.  An IRS  with total area $a \times b$ is placed on the $yz$-plane to assist the communication between the BS and UE.  Both the BS and UE have two antennas whereas the IRS is equipped with $N_y \times N_z = N$ elements.

 The first BS antenna (the one that is closest to the origin) is located at $(x_b,y_b,z_b)$ and the first antenna of UE is at $(x_u, y_u, z_u) $.  The location of each antenna can be written in three dimensions as follows.  The location of the $(m,n)^{th}$ element at IRS is
\begin{equation}
(0,\ (m-1) d^r_H \lambda_c, \ h + (n-1) d^r_H \lambda_c),
\end{equation}
where $m \in \left\lbrace 1,\dots, N_y\right\rbrace $, $ n \in \left\lbrace 1,\dots, N_z\right\rbrace $, $h$ is the height, $\lambda_c$ is the carrier frequency, and $d^r_H$ is the length of a square IRS element in wavelengths.
Similarly, the location of the $l^{th}$ antenna at UE is $
(x_u + (l-1) d^u_H \lambda_c \sin\theta_r \sin \varphi_r,  \ y_u + (l-1) d^u_H \lambda_c \sin\theta_r \cos \varphi_r, \ z_u + (l-1) d^u_H \lambda_c\cos\theta_r)$
where $l \in \left\lbrace 1,2 \right\rbrace $ and $d^u_H$ is the antenna spacing of the uniform linear array (ULA) at UE. The parameters  $\varphi_r$ and $\theta_r$  denote the azimuth and elevation angles, respectively, in local spherical coordinates at UE.

The location of antenna $s$ at BS is $
(x_b + (s-1) d^b_H \lambda_c \sin\theta_t \sin \varphi_t, \\y_b + (s-1) d^b_H \lambda_c \sin\theta_t \cos \varphi_t, \ z_b + (s-1) d^b_H \lambda_c \cos\theta_t)$
where $s \in \left\lbrace 1,2 \right\rbrace $ and $d^b_H$ is the antenna spacing of the ULA at BS. The parameters $\varphi_t$ and $\theta_t$  denote the azimuth and elevation angles, respectively, in local spherical coordinates at BS.

The BS is assumed to be in the far-field of the IRS and the channel between them is denoted with $\mathbf{H}_\mathrm{br} \in \mathbb{C}^{N\times 2}$.   We define the distance between the first elements of the IRS and BS as $d_\mathrm{br} = ( x^2_b  + y^2_b + (z_b - h)^2)^{1/2}$. In the far-field case, the antenna array lengths are negligible compared to the propagation distance, i.e., $d_\mathrm{br} \gg 2 d^u_H \lambda_c$ and $d_\mathrm{br} \gg \max(N_y,N_z) d^r_H \lambda_c$. Then, we write down the distances between each antenna pair and use the Maclaurin series expansion $(1 + \Delta)^{1/2} \approx 1 +\frac{\Delta}{2}$ to obtain
\begin{align}
&d_{m,n,s} \approx  d_\mathrm{br} + \lambda_c \Omega^{br}_{m,n,s}(\theta_t, \varphi_t),
\end{align}
where $\Omega^\mathrm{br}_{m,n,s} =  \Omega^\mathrm{br}_{i,s}= \frac{1}{d_\mathrm{br}} ((s-1) x_b   d^b_H  \sin\theta_t \sin\varphi_t 
+  y_b ((s-1) d^b_H  \sin\theta_t \cos\varphi_t - (m-1) d^r_H  ) +    (z_b - h)( (s-1) d^b_H \cos\theta_t - (n-1) d^r_H ) )$ where $ i = (m-1) N_z + n \in \left\lbrace 1, \dots,N\right\rbrace $. Note that $\Omega^\mathrm{br}_{1,1}=0$ and  $d_{111} = d_{br}$. The normalized channel  between the BS and IRS becomes 
\[\mathbf{H}_{\mathrm{br}} =   e^{j \frac{2 \pi d_\mathrm{br}}{\lambda_c} }
\begin{bmatrix}
 e^{j 2 \pi\Omega^\mathrm{br}_{1,1}}  &e^{j 2 \pi \Omega^\mathrm{br}_{1,2} }\\
 \vdots & \vdots\\
e^{j  2 \pi\Omega^\mathrm{br}_{N,1}}& e^{j 2 \pi \Omega^\mathrm{br}_{N,2} }
\end{bmatrix}.
\]
Similarly, the normalized channel between the IRS and UE is denoted by $\mathbf{H}_\mathrm{ru} \in  \mathbb{C}^{2 \times N}$  and has entries $ [\mathbf{H}_\mathrm{ru}]_{l,i} = e^{j \frac{2 \pi d_\mathrm{ru}}{\lambda_c} } e^{j 2 \pi \Omega^\mathrm{ru}_{l,i}}$, where 
\begin{align}
\Omega^\mathrm{ru}_{l,i} = & \frac{1}{d_\mathrm{ru}} \left[ (l-1) x_u   d^u_H  \sin\theta_r \sin\varphi_r \nonumber \right. \\
&+  y_u ((l-1) d^u_H  \sin\theta_r \cos\varphi_r - (m-1) d^r_H  ) \nonumber  \\
& \left.+    (z_u - h)( (l-1) d^u_H \cos\theta_r - (n-1) d^r_H ) \right]
\end{align}
and $d_\mathrm{ru} = ( x^2_u  + y^2_u + (z_u - h)^2)^{1/2}$. For the direct channel between the BS and UE, we write the steering vectors as
\begin{align}
\mathbf{a}_\mathrm{bs}= \begin{bmatrix}
1   \\
e^{j {2 \pi}  \Omega_\mathrm{bs}} 
\end{bmatrix}, 	\quad
&\mathbf{a}_\mathrm{ue} =  \begin{bmatrix}
1   \\
e^{j {2 \pi}  \Omega_\mathrm{ue}}
\end{bmatrix},
\end{align}
where
\begin{align}
	\Omega_\mathrm{bs} &= \frac{1}{d_\mathrm{bu}}\left[ (x_b - x_u)  d^b_H \sin\theta_t \sin\varphi_t   \right.  \\ 
	&+ (y_b - y_u) d^b_H \sin\theta_t \cos\varphi_t   + (z_b - z_u) d^b_H \cos\theta_t  \left.\right], \nonumber
\end{align}
\begin{align}
\Omega_\mathrm{ue}& = \frac{1}{d_\mathrm{bu}}\left[ (x_b - x_u)  d^u_H \sin\theta_r \sin\varphi_r   \right.  \\ 
&+ (y_b - y_u) d^u_H \sin\theta_r \cos\varphi_r   + (z_b - z_u) d^u_H \cos\theta_r  \left.\right], \nonumber
\end{align}
where the distance between BS and UE is $d_\mathrm{bu} =( (x_b- x_u)^2  + (y_b -y_u)^2 + (z_b - z_u)^2)^{1/2}$.
 Then, the channel is $\mathbf{H}_\mathrm{bu} =  \sqrt{\beta_\mathrm{bu}}e^{j\phi_\mathrm{bu}} \mathbf{a}_\mathrm{ue}\mathbf{a}^H_\mathrm{bs} $ where $\beta_\mathrm{bu}$ is the direct channel pathloss and $e^{j\phi_\mathrm{bu}} =e^{j \frac{2 \pi d_\mathrm{bu}}{\lambda_c} } $.

\section{Downlink Transmission}
In this section, we compare two cases: direct and IRS-aided downlink transmission. We will thereby demonstrate the channel rank improvement ability of the IRS.
\subsection{Case 1: Direct Transmission}

In this setup, the BS directly sends a signal to the UE without an assisting IRS. The received signal $\mathbf{y}_1 \in \mathbb{C}^{2\times 1}$ is
\begin{equation}
\mathbf{y}_1 =   \mathbf{H}_\mathrm{bu} \mathbf{V}_1 \mathbf{x} +\mathbf{n},
\end{equation}
where $\mathbf{x}\in \mathbb{C}^{2 \times 1}$ is the transmitted signal with power allocation matrix $\mathbf{P} $, $\mathbf{V}_1 \in \mathbb{C}^{2 \times 2}$ is the  downlink pre-processing matrix, and $\mathbf{n} \sim \mathcal{CN}(0, \sigma^2\mathbf{I})$ is  AWGN. The matrix $\mathbf{H}_\mathrm{bu} $ has singular value decomposition (SVD) $\mathbf{H}_\mathrm{bu} = \mathbf{U}_1 \boldsymbol{\Lambda}_1 \mathbf{V}_1^H$ where $\mathbf{U}_1 \in \mathbb{C}^{2\times 2}$, $\mathbf{V}_1$ are unitary matrices and $\boldsymbol{\Lambda}_1 \in \mathbb{R}^{2\times 2}$ is the diagonal singular value matrix. Then, the processed received signal at the UE is written as 
\begin{align}
\mathbf{U}_1^H\mathbf{y}_1 &=   \mathbf{U}_1^H\mathbf{H}_\mathrm{bu} \mathbf{V}_1 \mathbf{x} +\tilde{\mathbf{n}}_1 \nonumber \\
& =\boldsymbol{\Lambda}_1 \mathbf{x} +\tilde{\mathbf{n}}_1,
\end{align}
where $\tilde{\mathbf{n}}_1 = \mathbf{U}_1^H \mathbf{n}$.
The singular values of $\mathbf{H}_\mathrm{bu}$ are $\lambda^d_1 = 2 \sqrt{\beta_\mathrm{bu}}$ and $\lambda^d_2 =0 $. Therefore, the matrix is rank-deficient and can only support a  single data stream. Then, the BS transmits the signal and allocates all available power ${P}_\mathrm{tot}$ along the first right singular-vector of $\mathbf{H}_\mathrm{bu}$ i.e., $\mathbf{P} = \mathrm{diag}\left( P_\mathrm{tot}, 0\right) $. The direct channel capacity is \cite{Telatar1999}
\begin{equation}
R_1= \log_2 \left( 1 + \frac{P_\mathrm{tot} (\lambda^d_1)^2 }{\sigma^2} \right).
\end{equation}

\subsection{Case 2: IRS-aided transmission}

In this case, we assume that an IRS assists the communication through BS and UE. The received signal $\mathbf{y}_2 \in \mathbb{C}^{2\times 1}$ is
\begin{equation}
\mathbf{y}_2 =  \left( \mathbf{H}_\mathrm{ru} \boldsymbol{\Phi}\mathbf{H}_\mathrm{br} + \mathbf{H}_\mathrm{bu}\right) \mathbf{V}_2 \mathbf{x} +\mathbf{n}, 
\end{equation}
where $\mathbf{V}_2$ is the downlink pre-processing matrix, $\boldsymbol{\Phi} = \alpha \mathrm{diag}(e^{j\phi_1},\\ e^{j\phi_2},  \dots,e^{j\phi_{N}}) \in \mathbb{C}^{N \times N}$ is the local phase matrix with  phase coefficients  $\phi_1,\phi_2 \dots,\phi_{N}$ at each surface element. We assume that the scattering amplitude coefficient is $\alpha=1$.

 In order to support multiple data streams, the channel $\mathbf{H} = \mathbf{H}_\mathrm{ru} \boldsymbol{\Phi}\mathbf{H}_\mathrm{br} + \mathbf{H}_\mathrm{bu} $ should have $\mathrm{rank}\left(  \mathbf{H}\right) = 2$ and a good condition number $\frac{\lambda_1}{\lambda_2}$ where $\lambda_1$ and $\lambda_2$ are the singular values of $\mathbf{H}$. As mentioned in the previous section, the direct channel has $\mathrm{rank}\left( \mathbf{H}_\mathrm{bu} \right) = 1$.  The compound channel
$\mathbf{H}_\mathrm{c} =\mathbf{H}_\mathrm{ru} \boldsymbol{\Phi}\mathbf{H}_\mathrm{br} \in \mathbb{C}^{2 \times 2}$ is
\begin{align}
		&\mathbf{H}_\mathrm{c} = \sqrt{\beta_\mathrm{c}} e^{j\phi_c} \nonumber\\
		& \times   \begin{bmatrix}
		\displaystyle\sum_{i=1}^{N} e^{j\left( \phi_{i} +  2 \pi \Omega^\mathrm{ru}_{1,i}  +  2 \pi \Omega^\mathrm{br}_{i,1}   \right)}    &\displaystyle\sum_{i=1}^{N} e^{j\left( \phi_{i} + 2 \pi \Omega^\mathrm{ru}_{1,i}   + 2 \pi \Omega^\mathrm{br}_{i,2}   \right)}  \\
		\displaystyle\sum_{i=1}^{N} e^{j\left( \phi_{i} +  2 \pi \Omega^\mathrm{ru}_{2,i} +  2 \pi \Omega^\mathrm{br}_{i,1}    \right)}  & \displaystyle\sum_{i=1}^{N} e^{j\left( \phi_{i} +  2 \pi \Omega^\mathrm{ru}_{2,i}  +  2 \pi \Omega^\mathrm{br}_{i,2}   \right)}  
		\end{bmatrix},
\end{align}
where $\beta_\mathrm{c} $ is the pathloss of the scattered path and $e^{j\phi_c} = e^{j \frac{2 \pi }{\lambda_c}(d_\mathrm{br} +d_\mathrm{ru}) }$. The total channel can be decomposed as $\mathbf{H} = \mathbf{U}_2 \boldsymbol{\Lambda}_2 \mathbf{V}_2^H$. All matrices (i.e., the pre-processing matrix $\mathbf{V}_2$, post-processing matrix $\mathbf{U}_2$ and $\boldsymbol{\Lambda}_2$) depend on the phase matrix $\boldsymbol{\Phi}$. For any given $\boldsymbol{\Phi}$, we can write the processed signal at UE as
\begin{equation}
	\mathbf{U}_2^H \mathbf{y}_2 = \boldsymbol{\Lambda}_2  \mathbf{x} + \tilde{\mathbf{n}}_2,
\end{equation}
where $\tilde{\mathbf{n}}_2 = \mathbf{U}_2^H \mathbf{n}$. The singular value matrix  $ \boldsymbol{\Lambda}_2$ is not a function of  $\mathbf{P}= \mathrm{diag}(P_1, P_2)$. We can write the rate as a function of the local phases as
\begin{align}\label{eq:rate}
	R_2\left( \phi_1, \dots,\phi_N \right) = \sum_{j=1}^{2}\log_2 \left( 1 + \frac{P_j \lambda^2_j\left( \phi_1, \dots,\phi_N \right)  }{\sigma^2} \right).
\end{align}
 For any singular values, the optimal power allocation between the data streams is accomplished by using the water filling algorithm as
\begin{equation}\label{eq:water}
P_j =  \left(  \mu - \frac{\sigma^2}{\lambda^2_j} \right)^+, 
\end{equation}
where $\mu $ is chosen to satisfy the total power constraint $P_1 +P_2 = P_\mathrm{tot}$. The eigenvalues of $\mathbf{H}\mathbf{H}^H$, $\lambda^2_1$ and $\lambda^2_2$, are in the form of roots of a quadratic function i.e.,  $\lambda^2_1 = \frac{- b + \sqrt{b^2 -4c}}{2}$ and $\lambda^2_2 = \frac{- b - \sqrt{b^2 -4c}}{2}$ where  
\begin{align}
b = & - \left(|[\mathbf{H}]_{11}|^2 + |[\mathbf{H}]_{12}|^2 +|[\mathbf{H}]_{21}|^2 +|[\mathbf{H}]_{22}|^2\right), \\
c = & |[\mathbf{H}]_{11}|^2|[\mathbf{H}]_{22}|^2 + |[\mathbf{H}]_{12}|^2|[\mathbf{H}]_{21}|^2 \nonumber\\
&- 2 \mathrm{Re}\left\lbrace  [\mathbf{H}]_{11}[\mathbf{H}]^*_{12}[\mathbf{H}]^*_{21}[\mathbf{H}]_{22}\right\rbrace    
\end{align}
are functions of the local surface phases. Using the waterfilling algorithm with $\mu = \frac{1}{2}\left(P_\mathrm{tot} - \frac{\sigma^2 b}{c} \right) $, we can write the rate as $	R_2 = \log_2\left( \left( P_\mathrm{tot} - \frac{\sigma^2 b}{c}\right)^2  \frac{c}{4 \sigma^4} \right)$. In the high SNR regime, $R_2 \approx \log_2\left( \frac{P^2_\mathrm{tot}c - 2 \sigma^2 P_\mathrm{tot} b}{4 \sigma^4} \right)$ and it is maximized by  the selection of  $\phi_{i}$ that maximizes
\begin{align}\label{eq:structure}
&\!\!\!\!\! \left| \sqrt{\beta_\mathrm{c}} e^{j\phi_\mathrm{c}} \sum_{i=1}^{N} e^{j\left( \phi_{i} +  2 \pi \Omega^\mathrm{ru}_{1,i}  +  2 \pi \Omega^\mathrm{br}_{i,1}   \right)} + \sqrt{\beta_\mathrm{bu}}e^{j\phi_\mathrm{bu}}\right|^2 \nonumber\\
&\!\!\!\!\!\!\times \left| \sqrt{\beta_\mathrm{c}} e^{j\phi_\mathrm{c}} \sum_{i=1}^{N} e^{j\left( \phi_{i} +  2 \pi \Omega^\mathrm{ru}_{2,i}  +  2 \pi \Omega^\mathrm{br}_{i,2}   \right)} + \sqrt{\beta_\mathrm{bu}}e^{j\phi_\mathrm{bu}} e^{j {2 \pi}  (\Omega_\mathrm{ue}- \Omega_\mathrm{bs})}  \right|^2 \nonumber\\
&\!\!\!\! \!\!+ \left| \sqrt{\beta_\mathrm{c}} e^{j\phi_\mathrm{c}}  \sum_{i=1}^{N} e^{j\left( \phi_{i} +  2 \pi \Omega^\mathrm{ru}_{1,i}  +  2 \pi \Omega^\mathrm{br}_{i,2}   \right)} + \sqrt{\beta_\mathrm{bu}}e^{j\phi_\mathrm{bu}} e^{-j {2 \pi}  \Omega_\mathrm{bs}}\right|^2  \times \nonumber\\
&\!\!\!\!\!\!\left| \sqrt{\beta_\mathrm{c}} e^{j\phi_\mathrm{c}} \sum_{i=1}^{N} e^{j\left( \phi_{i} +  2 \pi \Omega^\mathrm{ru}_{2,i}  +  2 \pi \Omega^\mathrm{br}_{i,1}   \right)} + \sqrt{\beta_\mathrm{bu}}e^{j\phi_\mathrm{bu}}  e^{j {2 \pi}  \Omega_\mathrm{ue}}\right|^2\!.
\end{align}
From \eqref{eq:structure}, we select $\phi_i$ at each surface element that maximizes $   \cos(2 \phi_{i} +  2 \pi ( \Omega^\mathrm{ru}_{1,i}  + \Omega^\mathrm{br}_{i,1} + \Omega^\mathrm{ru}_{2,i}  + \Omega^\mathrm{br}_{i,2} +  \Omega_\mathrm{bs}-\Omega_\mathrm{ue})  ) $ that is 
\begin{equation}\label{eq:phase}
\phi^*_i = -\pi\left( \Omega^\mathrm{ru}_{1,i}  +  \Omega^\mathrm{br}_{i,1} + \Omega^\mathrm{ru}_{2,i}  +  \Omega^\mathrm{br}_{i,2} + \Omega_\mathrm{bs} - \Omega_\mathrm{ue}\right).
\end{equation}

\subsection{Deployment Analysis}
In the high SNR regime, inserting $\phi^*_i$ from \eqref{eq:phase} and $P_j$ from \eqref{eq:water} into the rate expression \eqref{eq:rate} gives
\begin{align}\label{eq:opt}
&	R_2\left( \phi^*_1, \dots,\phi^*_N \right)  = \sum_{j=1}^{2}\log_2 \left( 1 + \frac{P_j \lambda^2_j\left( \phi^*_1, \dots,\phi^*_N \right)  }{\sigma^2} \right) \nonumber \\
	&\approx \log_2\left( \frac{P^2_\mathrm{tot} \lambda^2_1(\phi^*_1, \dots,\phi^*_N) \lambda^2_2(\phi^*_1, \dots,\phi^*_N)}{4 \sigma^4}\right)  \nonumber \\
	&= \log_2\left(  \frac{P^2_\mathrm{tot} N^2 \beta_\mathrm{c} \beta_\mathrm{bu}}{\sigma^4}\Upsilon \right),
\end{align}
where $\Upsilon =	 \left(  1 - \cos\left( 2\pi (\Omega_\mathrm{br} +\Omega_\mathrm{bs})\right)  \right)\left( 1 - \cos\left( 2\pi (\Omega_\mathrm{ru} -\Omega_\mathrm{ue})\right) \right), $
where $ \Omega_\mathrm{ru} = \Omega^\mathrm{ru}_{2,i} - \Omega^\mathrm{ru}_{1,i} = \frac{1}{d_\mathrm{ru}} (  x_u   d^u_H  \sin\theta_r \sin\varphi_r 
+  y_u d^u_H  \sin\theta_r \\ \cos\varphi_r +    (z_u - h)d^u_H \cos\theta_r) $  and $
\Omega_\mathrm{br} = \Omega^\mathrm{br}_{i,2} -\Omega^\mathrm{br}_{i,1} = \frac{1}{d_\mathrm{br}} (x_b   d^b_H  \sin\theta_t \\ \sin\varphi_t 
+  y_b  d^b_H  \sin\theta_t \cos\varphi_t +    (z_b - h)d^b_H \cos\theta_t )$ do not depend on $i$. From \eqref{eq:opt}, we note that the SNR scales with $N^2$ and the rate is a function of the  BS, IRS and UE positions and their deployment angles. The arguments of the cosine functions in  $\Upsilon$ are
\begin{align}
	\Omega_\mathrm{br} + \Omega_\mathrm{bs} &= d^b_H \sin\theta_t  \sin\varphi_t\left( \frac{x_b}{d_\mathrm{br}} + \frac{(x_b -x_u )}{d_\mathrm{bu}}\right) \nonumber \\
	&+ d^b_H \sin\theta_t  \cos\varphi_t\left( \frac{y_b}{d_\mathrm{br}} + \frac{(y_b -y_u )}{d_\mathrm{bu}}\right) \nonumber \\
	&+ d^b_H \cos\theta_t  \left( \frac{(z_b - h)}{d_\mathrm{br}} + \frac{(z_b -z_u )}{d_\mathrm{bu}}\right), \\
\Omega_\mathrm{ru} - \Omega_\mathrm{ue} &=d^u_H \sin\theta_r  \sin\varphi_r\left( \frac{x_u}{d_\mathrm{ru}} - \frac{(x_b -x_u )}{d_\mathrm{bu}}\right) \nonumber \\
&+ d^u_H \sin\theta_r  \cos\varphi_r\left( \frac{y_u}{d_\mathrm{ru}} - \frac{(y_b -y_u )}{d_\mathrm{bu}}\right) \nonumber \\
&+ d^u_H \cos\theta_r  \left( \frac{(z_u - h)}{d_\mathrm{ru}} - \frac{(z_b -z_u )}{d_\mathrm{bu}}\right).
\end{align}
For different scenarios, the positions of the BS, IRS, and UE can be optimized according to given deployment requirements using these formulas. In the numerical results section, we assume that the BS and IRS have fixed locations and optimize the position of the UE.

\section{Numerical Results}
\begin{figure}[h]
	\includegraphics[scale=0.8]{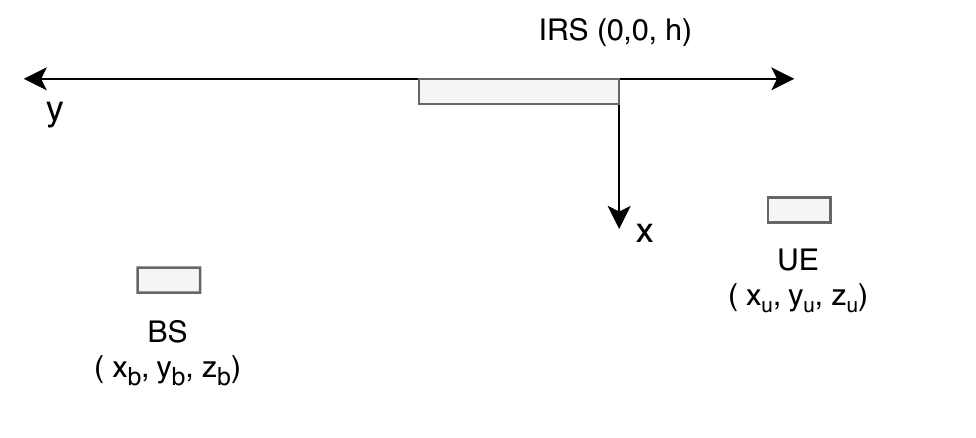}\vspace*{-6mm}
	\caption{Top view of simulation setup.}\label{simulationSetup4}
	\vspace*{-5mm}
\end{figure}

In this section, we quantify the capacity gains of adding  scattering paths to a MIMO system using an IRS. The simulation parameters are given in Table I and the simulation setup is illustrated in Fig.~\ref{simulationSetup4}. The pathloss of the direct path $\beta_\mathrm{bu}$ is calculated for $f_c = 5$ GHz using the model in \cite[Table B.1.2.1-1]{Umi} that is defined for $d_\mathrm{bu} \geq 10$ m.  In the simulations, we keep $N_z = 5 $ fixed and increase $N_y$ linearly.
The pathloss of the scattered path is calculated as \cite{Ben2019b}
\begin{align} \label{eq:ideal-pathloss}
\beta_\mathrm{c}= \frac{G_t G_r}{(4\pi)^2 } \left( \frac{a b}{d_\mathrm{br} d_\mathrm{ru}}\right) ^2  \cos^2(\varphi_i),
\end{align}
where  $a = N_z d^r_H \lambda_c$ and $b = N_y d^r_H \lambda_c$ are the surface dimensions and $\varphi_i = \mathrm{arctan}\left( \frac{y_b}{x_b}\right) $ is the angle of arrival to the surface. Note that $y_b \gg \max(a,b) $ and $x_b \gg \max(a,b) $.

\begin{table}[h!]
	\flushright
	\begin{tabular}{|l|l|lll}
		\cline{1-2}
		Parameter& Value     \\ \cline{1-2}
		Antenna and element spacings	& $d^b_H = d^u_H = 0.5$, $d^r_H =0.25 $     \\
		Location of BS $(x_b, y_b, z_b )$	&  $ (120, 120, 12)$ m   \\
		Location of UE $(x_u, y_u, z_u)$	&    $(5, -5, 1.5)$ m   \\
			Location of IRS &   $(0, 0, 2)$ m  \\
		Orientation angles of BS & $\theta_t =\pi/2 ,  \varphi_t  = 0$ \\
		Orientation angles of UE & $\theta_r =\pi/2 ,  \varphi_r  = 0$ \\
		Carrier frequency &  $f_c = 5$ GHz    \\
		Receiver noise power & $-94$ dBm \\
		Total power & $P_\mathrm{tot} = 10$ dBm \\
		Antenna gains at BS and UE & $G_t =G_r = 3$ dBi\\
		Pathloss of direct path, $\beta_\mathrm{bu}$ dB &  $ \!\! -41.97 -22 \log_{10}\left( d_\mathrm{bu}\right) + G_t + G_r \!\! $ \\\cline{1-2}
	\end{tabular}
	\caption{System parameters for the running example.}
\end{table}

\begin{figure}[t!]
	\includegraphics[scale=0.45]{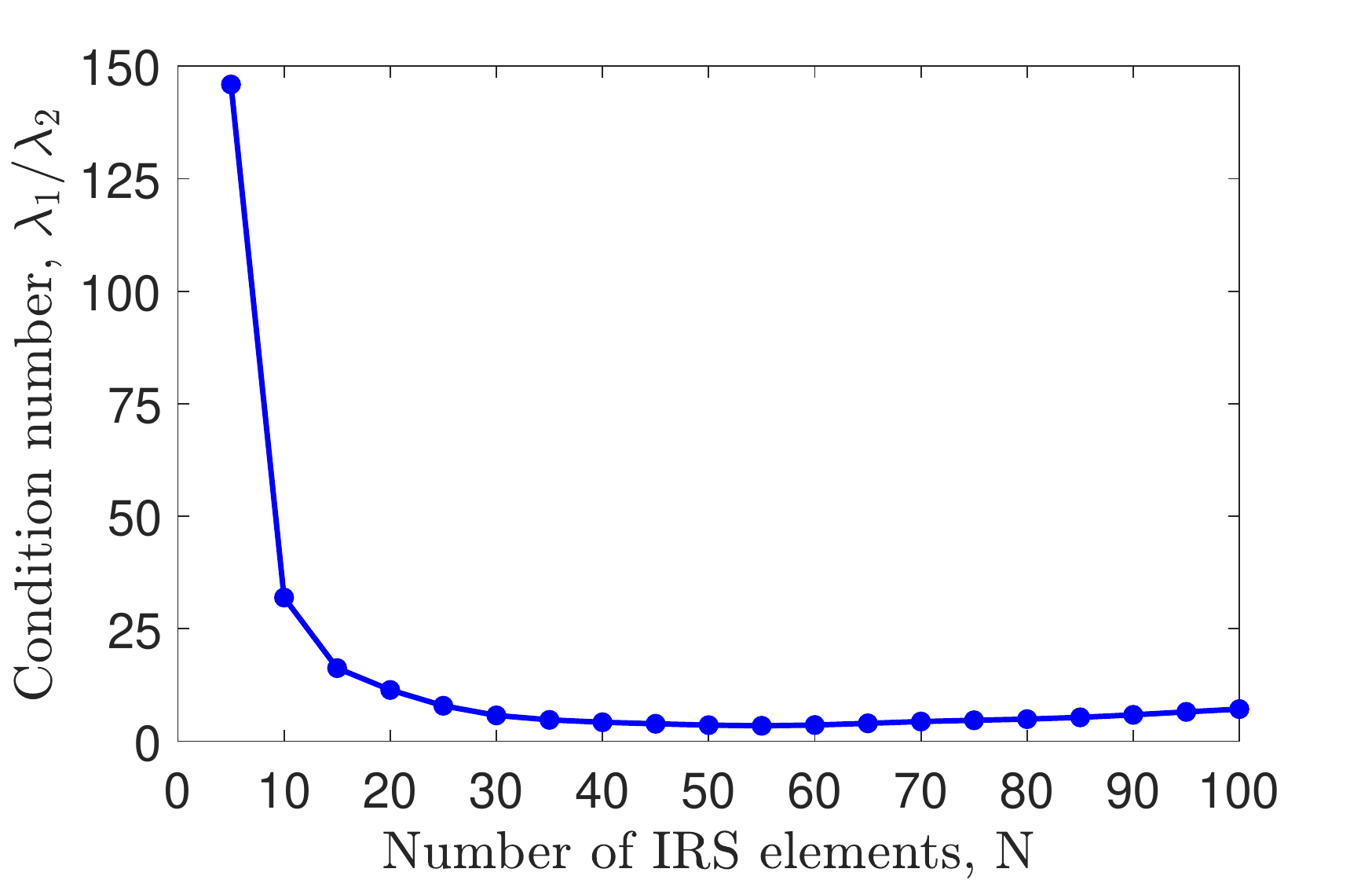}
	\caption{Condition number $\frac{\lambda_1}{\lambda_2}$ versus number of IRS elements.}\label{fig1}
	\vspace*{-4mm}
\end{figure}

Fig.~\ref{fig1} shows the condition number $\frac{\lambda_1}{\lambda_2}$ of the matrix $\mathbf{H}$ for different numbers of IRS elements. A matrix is said to be well-conditioned if the condition number is close to $1$ and such channel matrices support spatial multiplexing in the high SNR regime. As seen from the figure, $\frac{\lambda_1}{\lambda_2}\rightarrow\infty$ without an IRS (i.e., the matrix is rank deficient) while the ratio goes down as $N$ increases. 
After some point, it starts to increase again because $\lambda_1$ increases faster than $\lambda_2$ when the scattered path becomes stronger than the direct path.

\begin{figure}[t!]
	\includegraphics[scale=0.45]{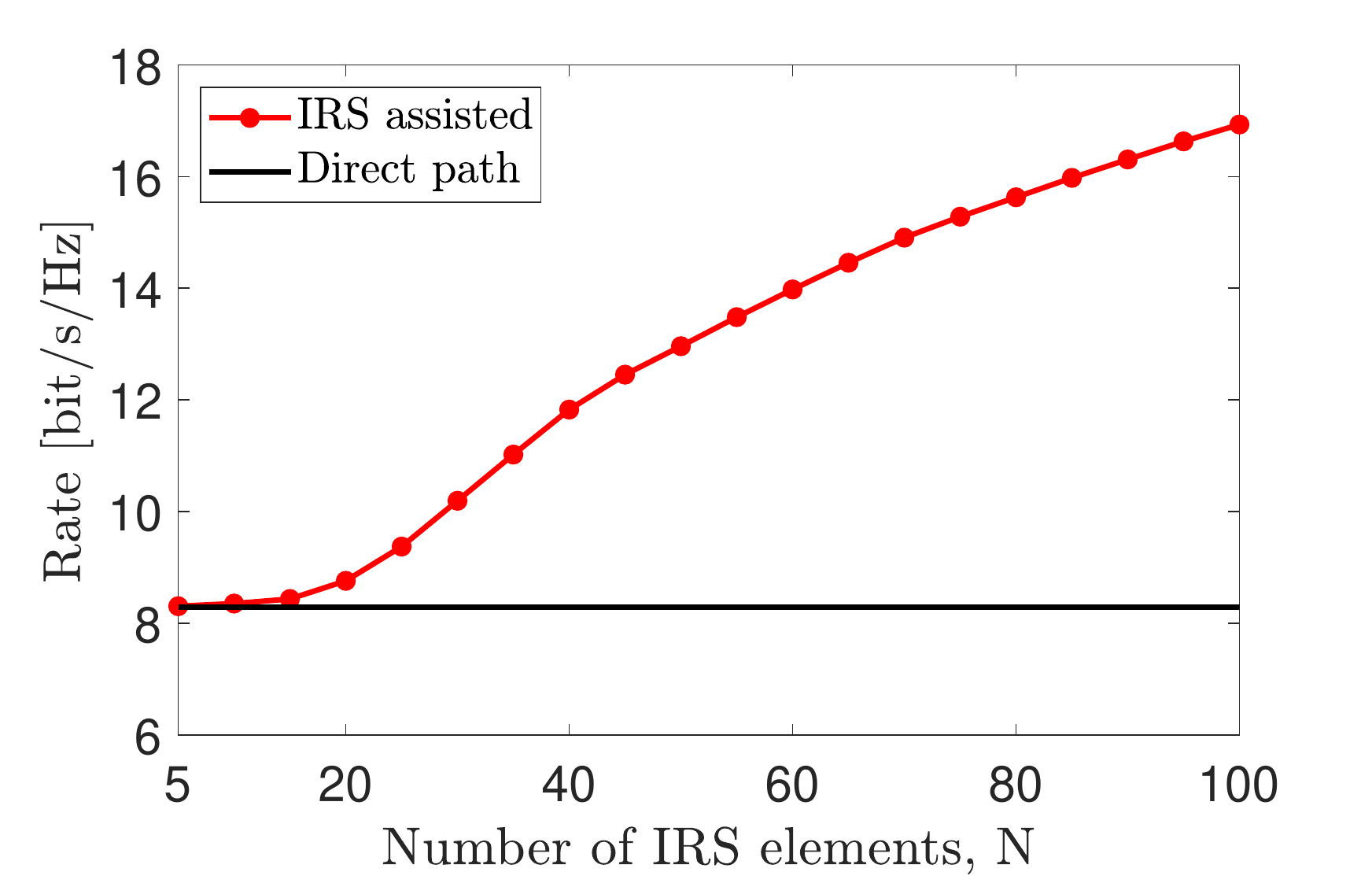}
	\caption{Rate versus number of IRS elements.}\label{fig2}
	\vspace*{-4mm}
\end{figure}

In Fig.~\ref{fig2}, we compare the rates of IRS-aided and direct transmissions. As expected, the rate increases with the number of IRS elements. Until $N=15$, the direct path dominates since it has smaller pathloss (i.e., larger SNR) but after $N=15$, we observe that the SNR in the IRS-aided case increases as $N^2$ and starts to outperform the direct transmission. The required $N$ to make the IRS practically useful highly depends on the locations of the BS and UE  (i.e., pathlosses). If the direct path is strong then the required value of $N$ for which the IRS becomes useful will also be high.

\begin{figure}[t!]
	\includegraphics[scale=0.45]{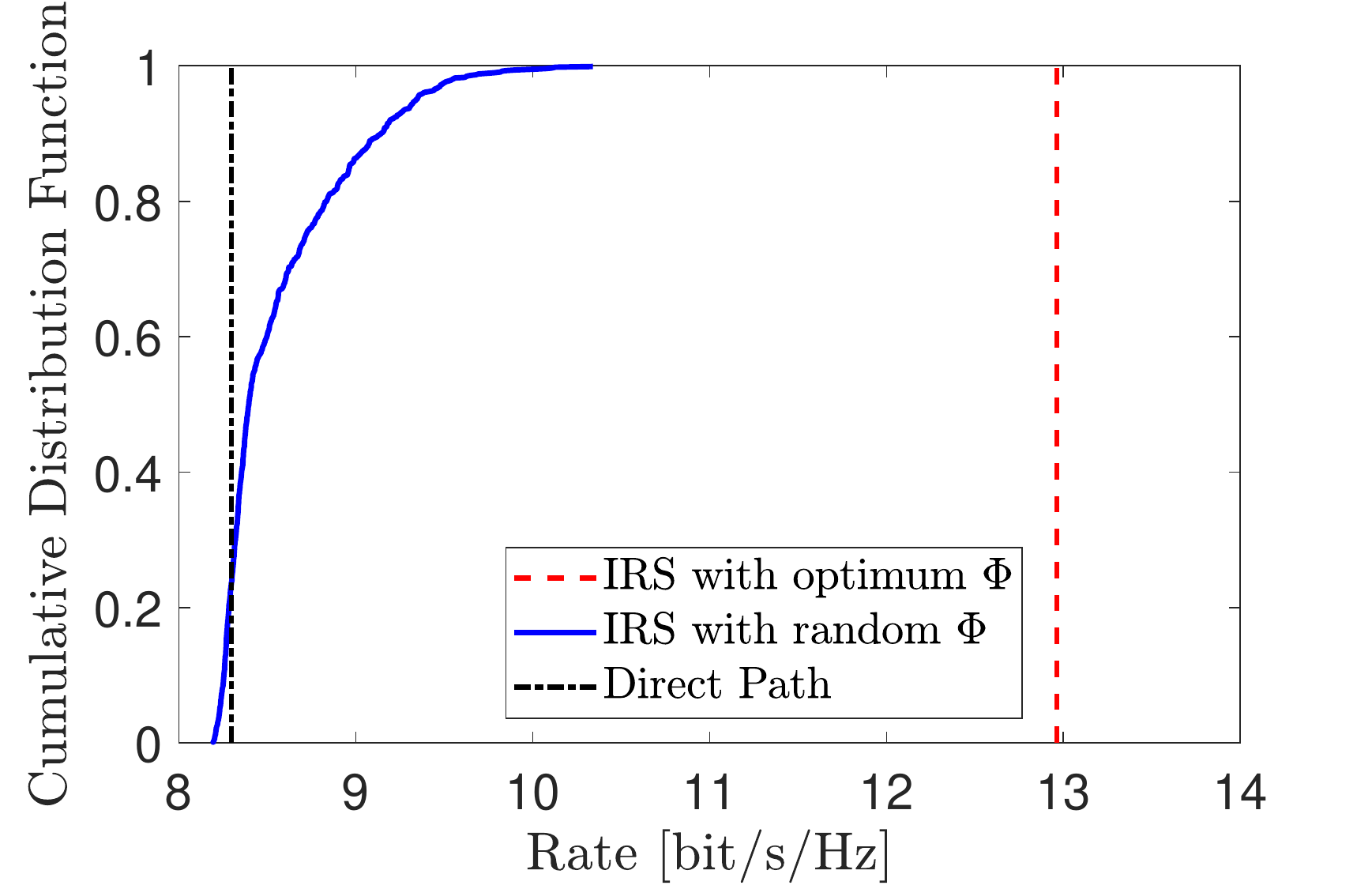}
	\caption{CDF of the rate for IRS with optimum and random phase matrix and direct transmission where $N = 50$.} \label{fig4}
	\vspace*{-4mm}
\end{figure}

Fig.~\ref{fig4} compares the rates of direct transmission and IRS-aided transmission with $N=50$ for the optimum or random selection of the local phase matrix $\boldsymbol{\Phi}$.  As seen from the figure, the phase matrix should be properly selected otherwise the phases at the UE may be destructively aligned, leading to a reduced rate when using the IRS.  As the number of $N$ increases, the performance gap between the random and optimized selection of phases will increase.

\begin{figure}[t!]
	\includegraphics[scale=0.45]{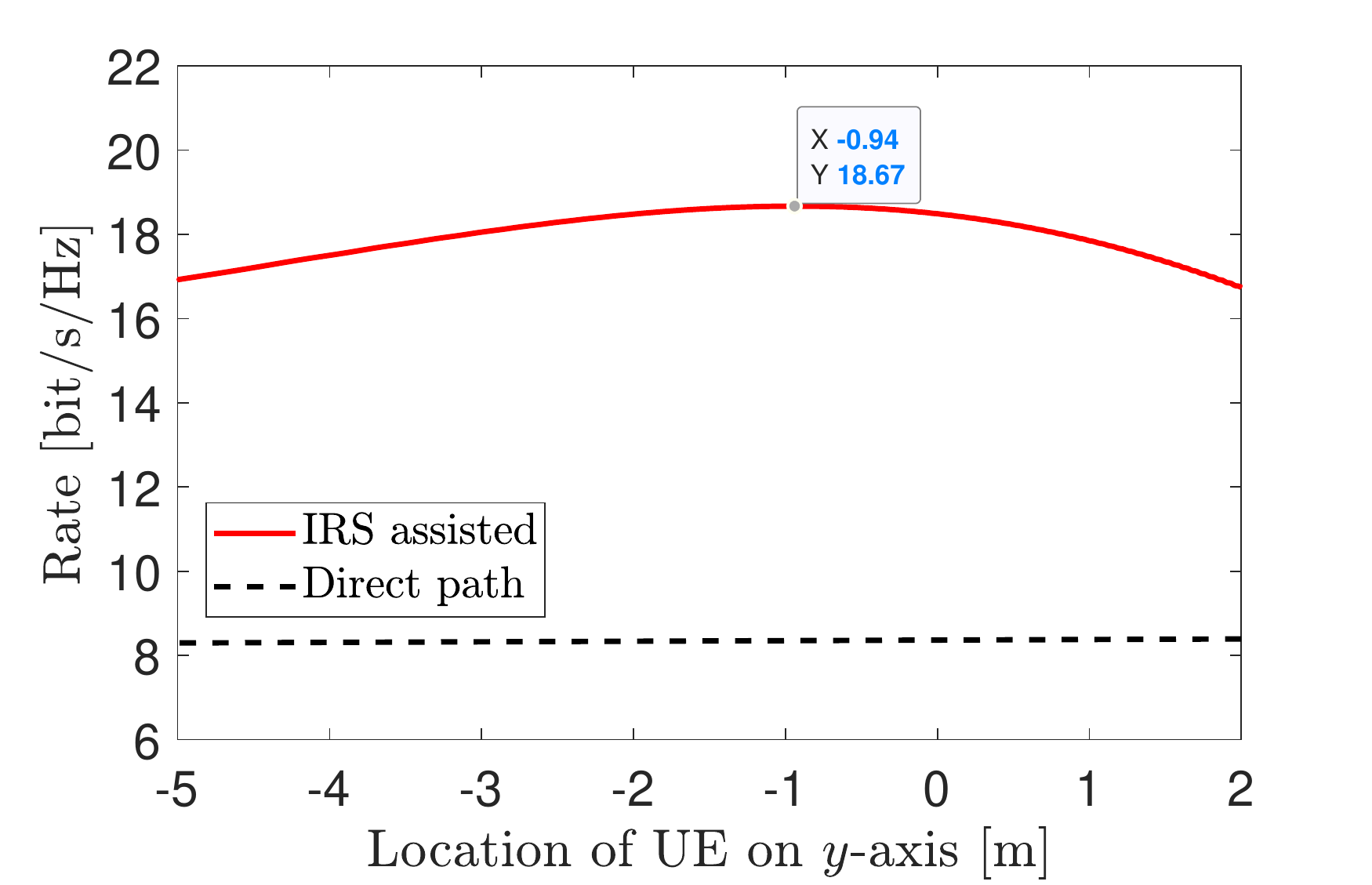}
	\caption{Rate versus location of UE on $y$-axis where $N=100$.} \label{fig6}
	\vspace*{-4mm}
\end{figure}

Fig.~\ref{fig6} shows the rate versus different UE locations on the $y$-axis. The locations of the BS and IRS are assumed fixed and the UE is moved from the point $y_u = -5$ to $y_u = 2$. By applying a linesearch algorithm, the maximum of the rate $\log_2\left( \frac{P^2_\mathrm{tot} N^2 \beta_\mathrm{c} \beta_\mathrm{bu}}{\sigma^4} \Upsilon \right)$ from \eqref{eq:opt} is obtained at  $y_u =- 0.94$. Inside the logarithm, there are three terms that are functions of the position $y_u$: the pathloss term $\beta_\mathrm{c} \beta_\mathrm{bu}$, and $\cos\left( 2\pi (\Omega_\mathrm{br} +\Omega_\mathrm{bs})\right)$, $\cos\left( 2\pi (\Omega_\mathrm{ru} -\Omega_\mathrm{ue})\right) $. The pathloss term $\beta_\mathrm{c} \beta_\mathrm{bu}$ has its maximum value around $y_u = 0$, however the cosine functions also affect the result and the maximum is obtained by the trade-off of these terms.


\section{Conclusion}
In this paper, we demonstrated the rank improvement ability of the recently emerged IRS technology. It enriches the propagation environment by adding multipaths with distinctively different spatial angles, so that a multiplexing gain is achieved even when the direct path has low rank. The performance greatly depends on the channel pathlosses and deployment angles. In order to reach its full potential, a careful deployment is necessary and the phases in the IRS must be properly selected, otherwise the rate can even be reduced.

\bibliographystyle{IEEEtran}
\bibliography{IEEEabrv,rankReferences}

\begin{thebibliography}{10}
\providecommand{\url}[1]{#1}
\csname url@samestyle\endcsname
\providecommand{\newblock}{\relax}
\providecommand{\bibinfo}[2]{#2}
\providecommand{\BIBentrySTDinterwordspacing}{\spaceskip=0pt\relax}
\providecommand{\BIBentryALTinterwordstretchfactor}{4}
\providecommand{\BIBentryALTinterwordspacing}{\spaceskip=\fontdimen2\font plus
\BIBentryALTinterwordstretchfactor\fontdimen3\font minus
  \fontdimen4\font\relax}
\providecommand{\BIBforeignlanguage}[2]{{%
\expandafter\ifx\csname l@#1\endcsname\relax
\typeout{** WARNING: IEEEtran.bst: No hyphenation pattern has been}%
\typeout{** loaded for the language `#1'. Using the pattern for}%
\typeout{** the default language instead.}%
\else
\language=\csname l@#1\endcsname
\fi
#2}}
\providecommand{\BIBdecl}{\relax}
\BIBdecl

\bibitem{Tretyakov}
V.~S. Asadchy, M.~Albooyeh, S.~N. Tcvetkova, A.~D\'{\i}az-Rubio, Y.~Radi, and
  S.~A. Tretyakov, ``Perfect control of reflection and refraction using
  spatially dispersive metasurfaces,'' \emph{Phys. Rev. B}, vol.~94, p. 075142,
  Aug 2016.

\bibitem{Alu}
N.~Mohammadi~Estakhri and A.~Al\`u, ``Wave-front transformation with gradient
  metasurfaces,'' \emph{Phys. Rev. X}, vol.~6, p. 041008, Oct 2016.

\bibitem{Yang2016}
H.~Yang, X.~Cao, F.~Yang, J.~Gao, S.~Xu, M.~Li, X.~Chen, Y.~Zhao, Y.~Zheng, and
  S.~Li, ``A programmable metasurface with dynamic polarization, scattering and
  focusing control,'' \emph{Scientific reports}, vol.~6, p. 35692, 2016.

\bibitem{Wan2016}
X.~Wan, M.~Q. Qi, T.~Y. Chen, and T.~J. Cui, ``Field-programmable beam
  reconfiguring based on digitally-controlled coding metasurface,''
  \emph{Scientific reports}, vol.~6, 2016.

\bibitem{Hum}
S.~V. {Hum} and J.~{Perruisseau-Carrier}, ``Reconfigurable reflectarrays and
  array lenses for dynamic antenna beam control: A review,'' \emph{IEEE
  Transactions on Antennas and Propagation}, vol.~62, no.~1, pp. 183--198, Jan
  2014.

\bibitem{Yu333}
N.~Yu, P.~Genevet, M.~A. Kats, F.~Aieta, J.-P. Tetienne, F.~Capasso, and
  Z.~Gaburro, ``Light propagation with phase discontinuities: Generalized laws
  of reflection and refraction,'' \emph{Science}, vol. 334, no. 6054, pp.
  333--337, 2011.

\bibitem{Wu2018}
Q.~{Wu} and R.~{Zhang}, ``Intelligent reflecting surface enhanced wireless
  network via joint active and passive beamforming,'' \emph{IEEE Transactions
  on Wireless Communications}, vol.~18, no.~11, pp. 5394--5409, Nov 2019.

\bibitem{Nadeem}
Q.-U.-A. Nadeem, A.~Kammoun, A.~Chaaban, M.~Debbah, and M.-S. Alouini,
  ``Asymptotic analysis of large intelligent surface assisted {MIMO}
  communication,'' \emph{CoRR}, vol. abs/1903.08127, 2019.

\bibitem{Akyildiz}
C.~{Liaskos}, S.~{Nie}, A.~{Tsioliaridou}, A.~{Pitsillides}, S.~{Ioannidis},
  and I.~{Akyildiz}, ``A new wireless communication paradigm through
  software-controlled metasurfaces,'' \emph{IEEE Communications Magazine},
  vol.~56, no.~9, pp. 162--169, Sep. 2018.

\bibitem{BJORNSON2019}
E.~Bj\"ornson, L.~Sanguinetti, H.~Wymeersch, J.~Hoydis, and T.~L. Marzetta",
  ``Massive {MIMO} is a reality—what is next?: Five promising research
  directions for antenna arrays,'' \emph{Digital Signal Processing}, 2019.

\bibitem{8741198}
C.~{Huang}, A.~{Zappone}, G.~C. {Alexandropoulos}, M.~{Debbah}, and C.~{Yuen},
  ``Reconfigurable intelligent surfaces for energy efficiency in wireless
  communication,'' \emph{IEEE Transactions on Wireless Communications},
  vol.~18, no.~8, pp. 4157--4170, Aug 2019.

\bibitem{Bjornson2019a}
E.~{Bj\"{o}rnson}, O.~{\"{O}zdogan}, and E.~G. {Larsson}, ``Intelligent
  reflecting surface vs. decode-and-forward: How large surfaces are needed to
  beat relaying?'' \emph{IEEE Wireless Communications Letters}, pp. 1--1, 2019.

\bibitem{Wu2019b}
Q.~{Wu} and R.~{Zhang}, ``Weighted sum power maximization for intelligent
  reflecting surface aided {SWIPT},'' \emph{IEEE Wireless Communications
  Letters}, pp. 1--1, 2019.

\bibitem{Guan}
X.~Guan, Q.~Wu, and R.~Zhang, ``Intelligent reflecting surface assisted secrecy
  communication via joint beamforming and jamming,'' \emph{arXiv preprint
  arXiv:1907.12839}, 2019.

\bibitem{Yu2019}
X.~Yu, D.~Xu, and R.~Schober, ``Enabling secure wireless communications via
  intelligent reflecting surfaces,'' \emph{arXiv preprint arXiv:1904.09573},
  2019.

\bibitem{Li2019}
S.~{Li}, B.~{Duo}, X.~{Yuan}, Y.~{Liang}, and M.~{Di Renzo}, ``Reconfigurable
  intelligent surface assisted {UAV} communication: Joint trajectory design and
  passive beamforming,'' \emph{IEEE Wireless Communications Letters}, pp. 1--1,
  2020.

\bibitem{Ye2019}
J.~Ye, S.~Guo, and M.-S. Alouini, ``Joint reflecting and precoding designs for
  {SER} minimization in reconfigurable intelligent surfaces assisted {MIMO}
  systems,'' \emph{arXiv preprint arXiv:1906.11466}, 2019.

\bibitem{Zhang2019a}
S.~Zhang and R.~Zhang, ``Capacity characterization for intelligent reflecting
  surface aided {MIMO} communication,'' \emph{arXiv preprint arXiv:1910.01573},
  2019.

\bibitem{1542408}
A.~{Lozano}, A.~M. {Tulino}, and S.~{Verdu}, ``High-{SNR} power offset in
  multiantenna communication,'' \emph{IEEE Transactions on Information Theory},
  vol.~51, no.~12, pp. 4134--4151, Dec 2005.

\bibitem{6509469}
E.~{Bj\"ornson}, M.~{Kountouris}, M.~{Bengtsson}, and B.~{Ottersten}, ``Receive
  combining vs. multi-stream multiplexing in downlink systems with
  multi-antenna users,'' \emph{IEEE Transactions on Signal Processing},
  vol.~61, no.~13, pp. 3431--3446, July 2013.

\bibitem{Telatar1999}
E.~Telatar, ``Capacity of multi-antenna gaussian channels,'' \emph{European
  transactions on telecommunications}, vol.~10, no.~6, pp. 585--595, 1999.

\bibitem{Umi}
``Further advancements for {E-UTRA} physical layer aspects (release 9).'' Mar
  2010.

\bibitem{Ben2019b}
O.~{\"{O}zdogan}, E.~{Bj\"{o}rnson}, and E.~G. {Larsson}, ``Intelligent
  reflecting surfaces: Physics, propagation, and pathloss modeling,''
  \emph{IEEE Wireless Communications Letters}, pp. 1--1, 2019.

\end{thebibliography}

\end{document}